\newcommand*{\ATLASLATEXPATH}{latex/}

\documentclass[cernpreprint,texlive=2016,txfonts,UKenglish]{latex/atlasdoc}

\pdfoutput=1
\usepackage[subfigure=true]{\ATLASLATEXPATH atlaspackage}
\usepackage{multirow}
\usepackage{float}
\usepackage{lscape}

\usepackage{\ATLASLATEXPATH atlasbiblatex}

\usepackage{\ATLASLATEXPATH atlascontribute}

\usepackage{\ATLASLATEXPATH atlasphysics}

\graphicspath{{logos/}{figures/}}

\nolinenumbers

\pdfoutput=1
\AtlasTitle{A search for resonances decaying into a Higgs boson and a new particle $X$ in the $XH \to qqbb$ final state with the ATLAS detector}

\author{The ATLAS Collaboration}

\AtlasRefCode{EXOT-2017-04}

\PreprintIdNumber{CERN-EP-2017-204} 

 \AtlasJournal{Phys.\ Lett.\ B.}
\AtlasDOI{10.1016/j.physletb.2018.01.042}

\AtlasAbstract{%
A search for heavy resonances decaying into a Higgs boson ($H$) and a new particle ($X$) is reported,
utilizing 36.1~\ifb\ of proton--proton collision data at $\sqrt{s} =$~13~\tev\ collected 
during 2015 and 2016 with the ATLAS detector at the CERN 
Large Hadron Collider. The particle $X$ is assumed to decay to a pair of light quarks, and
the fully hadronic final state $XH \rightarrow q\bar q'b\bar b$ is
analysed. The search considers the regime of high $XH$ resonance masses, 
where the $X$ and $H$ bosons are both highly Lorentz-boosted and are each reconstructed using a single jet with large radius parameter.
A two-dimensional phase space of $XH$ mass versus $X$ mass is scanned for evidence
of a signal, over a range of 
$XH$ resonance mass values between 1~\tev\ and 4~\tev, and for $X$ particles with masses from 50~\gev\ to 1000~\gev.
All search results are consistent with the expectations for the background due to Standard Model processes, and
95\% CL upper limits are set, as a function of $XH$ and $X$ masses, on the production cross-section of the $XH \to q\bar q'b\bar b$ resonance.
}

\hypersetup{pdftitle={ATLAS document},pdfauthor={The ATLAS Collaboration}}

\DeclareOldFontCommand{\rm}{\normalfont\rmfamily}{\mathrm}
\DeclareOldFontCommand{\sf}{\normalfont\sffamily}{\mathsf}
\DeclareOldFontCommand{\tt}{\normalfont\ttfamily}{\mathtt}
\DeclareOldFontCommand{\bf}{\normalfont\bfseries}{\mathbf}
\DeclareOldFontCommand{\it}{\normalfont\itshape}{\mathit}
\DeclareOldFontCommand{\sl}{\normalfont\slshape}{\@nomath\sl}
\DeclareOldFontCommand{\sc}{\normalfont\scshape}{\@nomath\sc}

\begin{document}
\nolinenumbers

\maketitle
\clearpage

%-------------------------------------------------------------------------------
\section{Introduction}
\label{sec:intro}
 
 The Standard Model (SM), including the recently discovered Higgs boson~\cite{HIGG-2012-27,CMS-HIG-12-028}, 
 describes collider phenomenology for energy scales up to a few hundred \gev. However, a number of issues
 with the SM, including sensitivity of the Higgs boson mass to radiative corrections, indicate either extreme fine-tuning
or the presence of new physics at an energy scale not far above the Higgs boson mass.
Many theoretical extensions to the SM predict the existence of new particles around the \tev\
scale, and it is natural to expect such particles to exhibit significant coupling to the Higgs boson ($H$).  
Examples of previous studies at the CERN Large Hadron Collider (LHC)
which are motivated by these considerations include ATLAS~\cite{EXOT-2015-18,Aaboud:2017ahz}
and CMS~\cite{Sirunyan:2017wto,CMS-EXO-14-009,CMS-EXO-14-010,CMS-EXO-13-007}
searches for new resonances decaying into
$VH$, where $V$ denotes a $W$ or $Z$ boson, 
as well as ATLAS~\cite{EXOT-2015-11} and CMS~\cite{CMS-EXO-12-053,CMS-HIG-14-013} searches 
for resonances decaying into $HH$. 

This Letter presents a new and more general search for heavy resonances which decay into the
SM Higgs boson and a new particle ($X$), utilizing a data sample 
corresponding to 36.1~\ifb\ of proton--proton ($pp$) collisions 
at $\sqrt{s} = $~13~\tev\, collected during 2015 and 2016 with the ATLAS detector. 
Since the hypothetical particle $X$ has not yet
been detected, it has unknown properties. It is
assumed that  $X$ decays hadronically into a pair of light quarks,
$X \to q\bar q'$, and has a natural width which is narrow compared 
to the detector resolution. 
It is also assumed that single $X$ 
production is sufficiently rare that it has not been observed by analyses such as the ATLAS and CMS searches for dijet resonances~\cite{Aaboud:2017yvp,CMS-EXO-16-032,Sirunyan:2017nvi}.

Examples of theories of physics beyond the SM which predict the existence of $XH$ 
resonances include pseudo Nambu--Goldstone 
models~\cite{Galloway:2016fuo}, extended Higgs sectors~\cite{Ellwanger:2017skc} and 
extra-dimension frameworks with warped 
compactifications~\cite{Agashe:2016kfr}.
For simplicity, such $XH$ resonances are hereafter denoted by $Y$. 
It is assumed that the new resonance $Y$ is narrow.
This study analyses the decay chain $Y \to XH \to q\bar q'b\bar b$, 
and considers the regime of high $Y$ masses, in the range 
between 1~\tev\ and 4~\tev, with $m(Y) \gg m(X)$,
which implies that
 the $X$ and $H$ bosons are both highly Lorentz-boosted. The $X$ and $H$ boson candidates are each reconstructed in a single
jet with large radius parameter; these jets are called 
``large-$R$ jets'' throughout this Letter and are denoted by $J$. 
Jet substructure techniques and $b$-tagging are used 
to suppress the dominant background from multijet
events, to enhance the sensitivity to the dominant $H \to b \bar b$ decay mode,
and to distinguish between the two bosons.

Given their assumed decay modes,
both $X$ and $Y$ must be bosons. However, the search does not impose 
explicit requirements on their electric charge, spin or polarization states,
and utilizes a simple event selection that does not seek to exploit angular
correlations or other variables that would be sensitive to particular
spin states. The analysis utilizes the reconstructed $Y$ mass as the final discriminant,
searching  for an excess due to a resonance in the $Y \to XH \to JJ$ mass distribution, 
separately in a large number of overlapping ranges of $X$ mass.

A modified form of the benchmark heavy vector triplet (HVT) model~\cite{Pappadopulo:2014qza,deBlas:2012qp}
is used to provide a measure of the implications of the results of the search within
one particular theoretical framework. The HVT model provides a simplified phenomenological Lagrangian that 
predicts heavy spin-1 resonances which are produced via $q\bar{q}$ annihilation, and
which can have significant branching ratios for 
decays into boson pairs, with longitudinally polarized bosons. 
The search results are used to set limits on the production
cross-section $\sigma (pp \to Y \to XH \to q\bar q'b \bar b)$ within the HVT
theoretical framework, to provide an interpretation
in the case where both $X$ and $Y$ have spin 1. The search
overlaps with the $VH$ and $HH$ searches, when the $X$ mass window
extends around the $V$ and $H$ mass values, respectively, in which cases the 
results are found to be compatible with those analyses. However, the
current search considers a much wider range of $X$ masses, from 50~\gev\ to 1000~\gev.

%-------------------------------------------------------------------------------

%-------------------------------------------------------------------------------
\section{ATLAS detector}
\label{sec:detector}

The ATLAS detector~\cite{PERF-2007-01} is a general-purpose particle physics detector 
used to investigate a broad range of physics processes. ATLAS includes an inner detector (ID) 
surrounded by a superconducting solenoid, electromagnetic (EM) and hadronic calorimeters, 
and a muon spectrometer (MS). The ID consists of a silicon 
pixel detector, including the insertable B-layer~\cite{IBL} installed after Run 1 of the LHC, a silicon microstrip 
detector and a straw-tube tracker. The ID is immersed inside a $2$~T axial magnetic field from 
the solenoid and provides precision tracking of charged particles with 
pseudorapidity\footnote{ATLAS uses a right-handed coordinate system with its origin 
at the nominal interaction point (IP) in the centre of the detector and the $z$-axis along the beam pipe. 
The $x$-axis points from the IP to the centre of the LHC ring, and the $y$-axis points 
upwards. Cylindrical coordinates $(r,\phi)$ are used in the transverse plane, $\phi$ being 
the azimuthal angle around the beam pipe. The pseudorapidity is defined in terms of the 
polar angle $\theta$ as $\eta = -\ln \tan(\theta/2)$.  
Angular distance is measured in units of 
$\Delta R \equiv \sqrt{(\Delta\eta)^{2} + (\Delta\phi)^{2}}$.} 
$|\eta| < 2.5$. The straw-tube tracker also provides transition radiation measurements for electron identification. 
The calorimeter system consists of finely segmented sampling calorimeters using 
lead/liquid-argon for the detection of EM showers up to $|\eta| =$ 3.2, and copper or 
tungsten/liquid-argon for hadronic showers for 1.5 $< |\eta| <$ 4.9. In the central region 
($|\eta| <$ 1.7), a steel/scintillator hadronic calorimeter is used. Outside the calorimeters, 
the MS incorporates multiple layers of trigger and tracking chambers within a 
magnetic field produced by a system of superconducting toroids, enabling an independent 
precise measurement of muon track momenta for $|\eta| <$ 2.7. The ATLAS detector has 
a two-level trigger system. The first-level (L1) trigger is implemented in hardware and uses the 
calorimeter and muon systems to reduce the accepted event rate to 100~kHz. The L1 trigger
is followed by a software-based high-level trigger 
to reduce the accepted event rate to approximately 1 kHz for offline analysis~\cite{TRIG-2016-01}.

%-------------------------------------------------------------------------------

%-------------------------------------------------------------------------------
\section{Signal and background simulation}
\label{sec:samples}

Samples of simulated signal and background events are used to optimize the event selection 
and to estimate the background contributions from various SM processes. 

The $Y \to XH$ signal was simulated by modifying the  
HVT~\cite{Pappadopulo:2014qza,deBlas:2012qp} configuration for $W' \to WH$ decays, replacing
the $W'$ boson with a narrow resonance  $Y$, and the $W$ boson by $X$. 
The new $X$ boson was assumed to have
spin 1, a natural width of 2 \gev, and a branching ratio ($B$) of 100\% for  
decays $X \to u\bar{d}$. The properties of the Higgs boson were assumed to be as described in the SM,
with its mass set to 125.5 \gev\ and decays into $b\bar{b}$ and $c\bar{c}$ enabled in the simulation. Given the SM branching ratios and the relevant experimental performance factors, Higgs boson decays to $b\bar{b}$ dominate the signal efficiency. The signal 
samples were generated with {\sc Madgraph5}\_a{\sc MC@NLO} 2.2.2~\cite{Alwall:2014hca} 
interfaced to \PYTHIA 8.186~\cite{Sjostrand:2007gs} for parton shower and hadronization, 
with the NNPDF2.3 next-to-leading-order (NLO) parton distribution function (PDF) set~\cite{Ball:2012cx} and the 
ATLAS A14 set of tuned parameters (tune)~\cite{ATL-PHYS-PUB-2014-021} for the underlying event. 
Approximately 100 signal samples were generated, each with individual values of 
$Y$ and $X$ masses, chosen in a two-dimensional grid of values ranging from 
1~\tev\ to 4~\tev \ for $m(Y)$ and from 50~\gev\ to 1000~\gev\ for $m(X)$.
The requirements $m(Y) >$ 1~\tev\ and $m(Y) \gg m(X)$ were applied, to
focus on the kinematic regime where the $X$ and $H$ bosons are highly
Lorentz-boosted.

The dominant SM background is due to multijet processes and,  as described in Section~\ref{sec:bkgndestimate},
is evaluated using data-driven techniques. Some cross-checks were
performed using simulated multijet events generated with \PYTHIA 8.186, 
with the NNPDF2.3 NLO PDF and the ATLAS A14 tune.

Minor background contributions, which are estimated using simulated samples,
are due to $t\bar{t}$ production and to vector boson
production in association with jets ($V$+jets).
The $t\bar{t}$ samples were generated with \textsc{Powheg-Box} v2~\cite{Frixione:2002ik} 
with the CT10 PDF set~\cite{Lai:2010vv}, interfaced with \PYTHIA 6.428~\cite{Sjostrand:2006za} 
and the Perugia 2012 tune for the parton shower~\cite{Skands:2010ak} using the CTEQ6L1 
PDF set~\cite{Pumplin:2002vw}. The cross-section of the $t\bar{t}$ process was normalized to 
the result of a QCD calculation at next-to-next-leading order and next-to-next-leading logarithmic accuracy (NNLO+NNLL), as implemented in \textsc{Top++} 2.0~\cite{Czakon:2011xx}. 
The $V$+jets background samples were generated with \SHERPA 2.1.1~\cite{Gleisberg:2008ta} 
interfaced with the CT10 PDF set, including only hadronic $V$ decays. Matrix elements of up 
to four extra partons were calculated at leading order in QCD. The $V$+jets events 
are normalized to the NNLO cross-sections~\cite{Anastasiou:2003ds}.%Anastasiou:2003ds}. 

For all simulated samples except those produced using \SHERPA, \textsc{EvtGen} v1.2.0~\cite{Lange:2001uf} was used to model the properties of bottom and charm hadron decays. 
The effect of multiple $pp$ interactions in the same and neighbouring bunch crossings (pile-up) 
was included by overlaying minimum-bias events simulated with \PYTHIA 8.186 
on each generated event~\cite{ATL-PHYS-PUB-2015-019}. The detector response was simulated with a {\textsc GEANT4}~\cite{Agostinelli:2002hh} based framework~\cite{SOFT-2010-01}
and the events were processed with the same reconstruction software as that used for data.

%-------------------------------------------------------------------------------

%-------------------------------------------------------------------------------
\section{Event reconstruction}
\label{sec:objects}
The reconstruction and identification of hadronically decaying bosons is performed using jets. 
Large-$R$ jets are reconstructed with 
the anti-$k_t$ algorithm~\cite{antikt,Cacciari:2011ma} with radius parameter $R = 1.0$ from three-dimensional topological 
clusters of energy deposits in the calorimeter~\cite{PERF-2014-07}. The large-$R$ jets are required to have transverse momentum \pt\ > 250~\gev\ and $|\eta| <$~2.0. % and jet mass $m_J >$~50~\gev. 

 Track-jets are formed from charged-particle tracks with 
$\pt > 0.4$ \gev\ and $|\eta| < 2.5$ that satisfy a set of hit and impact parameter criteria to reduce 
the contamination from pile-up interactions, and are clustered using the anti-$k_t$ algorithm with 
$R$ = 0.2~\cite{ATL-PHYS-PUB-2015-035}. Only track-jets with \pt > 25 \gev\ that are constructed from 
at least two tracks are considered in this analysis.

To mitigate pile-up effects and 
 soft radiation, the large-$R$ jets are trimmed~\cite{Krohn:2009th}, and then are subsequently reclustered 
 into subjets using the $k_t$ algorithm~\cite{Catani:1993hr}. To compensate for the limited angular 
 resolution of the calorimeter, the mass of large-$R$ jets is computed using a combination of 
 calorimeter and tracking information~\cite{ATLAS-CONF-2016-035}. The combined jet mass is defined as

\[
m_{J} \equiv w_{\rm calo} \times m_J^\mathrm{calo} + w_{\rm track} \times \Big( m_J^\mathrm{track} \frac{\pt^\mathrm{calo}}{ \pt^\mathrm{track}} \Big),
\label{eq:combmassdef}
\]
where $m_J^\mathrm{calo}$ is the calorimeter-only estimate of the jet mass. The variable $m_J^\mathrm{track}$ is the jet mass 
estimated via tracks with \pt $> 0.4$ \gev\  associated with the large-$R$ jet using ``ghost association''~\cite{antikt}, 
and is scaled in the combined mass formula by the ratio of calorimeter to track \pt estimates in order
to account for the missing neutral-particle component in the track-jet. The ghost association technique relies on repeating the jet clustering process with the addition of measured tracks that have the same direction but infinitesimally small \pt, so that the jet properties remain unaffected. A track is associated with a jet if it is contained in the jet after this re-clustering procedure. The weighting factors $w_{\rm calo}$ and $w_{\rm track}$ depend on \pt\, satisfy $w_{\rm calo} + w_{\rm track} = 1$, and are used to optimize the combined mass resolution. To partially account for the energy carried by muons from semileptonic $b$-hadron decays, 
the four-momentum of the closest muon candidate satisfying ``Tight'' muon identification criteria~\cite{PERF-2015-10} 
with $\pt\ > 4$~\gev\ and $|\eta| < 2.5$ that is within $\Delta R  = 0.2$ of a track-jet that is
$b$-tagged is added to the calorimeter jet four-momentum~\cite{ATLAS-CONF-2016-039}. In this study, only large-$R$ jets with $m_J >$~50~\gev\ are considered for further analysis.

Identifying a large-$R$ jet as a hadronically decaying 
Higgs boson candidate is aided by using track-jets matched via ghost association to the 
large-$R$ jet~\cite{ATL-PHYS-PUB-2014-013}.  The identification of $b$-hadrons relies on a 
multivariate $b$-tagging algorithm~\cite{ATL-PHYS-PUB-2016-012} applied to a set of tracks in a region of
interest around each track-jet axis. The $b$-tagging requirements
result in an efficiency of 77\% for 
track-jets containing $b$-hadrons, and the misidentification rate is $\sim 2\%$ ($\sim 24\%$) for 
light-flavour (charm) jets. These were determined in a sample of simulated $t\bar{t}$ events. For simulated 
samples the $b$-tagging efficiencies are corrected, based on the jet \pt, to match those measured in data~\cite{PERF-2012-04}.

Identifying a large-$R$ jet as a hadronically decaying $X \to q\bar q'$ 
boson candidate is aided by using jet substructure techniques. The variable $D_2$ is defined\footnote{The angular exponent $\beta$, defined in Ref.~\cite{Larkoski:2014gra}, is set to unity.} as a ratio of two- and three-point energy correlation functions~\cite{Larkoski:2014gra,Larkoski:2015kga}, 
which are based on the energies and pairwise angular distances of particles within a jet. 
This variable is optimized~\cite{conf:2017-064} to distinguish between jets originating from a single parton and those 
from the two-body decay of a heavy particle. A detailed description of the optimization,
performed using simulated $V$ decays, can be found in 
Refs.~\cite{PERF-2015-03,ATL-PHYS-PUB-2015-033}. Studies of boosted $W$ boson decays~\cite{PERF-2015-03} demonstrate that the $D_2$ variable is well modeled, with good agreement observed between data and simulation.

``Loose'' electrons and muons are reconstructed and
identified as described in Refs.~\cite{ATLAS-CONF-2016-024,PERF-2015-10}.
These leptons have $\pt\ > 7$ \gev, $|\eta| < 2.5\ (2.47)$ for muons (electrons), 
$|d_0| / \sigma_{d_0} < 3\ (5)$ and $|z_0 \sin \theta| < 0.5$ mm, where $d_0$ is transverse 
impact parameter with respect to the beam line, $\sigma_{d_0}$ is the corresponding uncertainty, 
and $z_0$ is the distance between the longitudinal position of the track along the beam line at the 
point where $d_0$ is measured and the longitudinal position of the primary vertex. An isolation 
criterion is also applied. Specifically, 
within a cone of size $\Delta R = 0.2 \ (0.3)$ around an electron (muon), the scalar sum of transverse momenta of tracks 
divided by the lepton \pt\ is required to be less than a cut-off value,
chosen  to provide a constant efficiency of around 99\% as a function of \pt and $|\eta|$. 

%-------------------------------------------------------------------------------

%----------------------------------------------------------	---------------------
\section{Event selection}
\label{sec:selection}

Events are selected from the 2015 (2016) running period 
using a trigger that requires a single large-$R$ jet 
with \pt > 360 (420)~\gev. 
Selected events are required to have at least one 
primary vertex (PV) that has at least two associated tracks, each with 
transverse momentum \pT > 0.4\,\gev. For events with more than one PV candidate, 
the one with the largest track $\sum\pT^2$ is chosen as the hard-scatter PV.
 To select a hadronic final state, events with one or more 
 ``Loose'' charged leptons (electrons or muons) are vetoed. 

Events are kept only if they contain at least two large-$R$ jets. For events with more than two large-$R$ jets, 
only the two ``leading'' jets, namely those with highest \pt, are considered 
as possible $X$ and $H$ candidates.  To ensure high and uniform trigger efficiency,
the leading large-$R$ jet must satisfy \pt > 450 \gev.

Additional requirements are imposed to distinguish hadronically decaying boson candidates 
from background jets. 
To be identified as a Higgs jet candidate, the combined mass of the jet is required to 
satisfy 75~\gev\ < $m_J$ < 145~\gev, a criterion that is $\sim$90\% efficient for Higgs 
boson jets. The number of $b$-tagged jets inside the Higgs jet is used to categorize the 
events in two orthogonal signal regions: the ``1-tag''  signal region (SR1)
and ``2-tag'' signal region (SR2) definitions
require Higgs candidates to contain exactly one and two $b$-tagged jets, respectively. The SR1 region
recovers some efficiency loss in the high $Y$ mass region where the angular separation between the 
two $b$-quarks from the Higgs decay is small and the fragmentation products are 
merged~\cite{ATLAS-CONF-2016-039}. As depicted in Fig.~\ref{fig:strategy} and
described in Section~\ref{sec:bkgndestimate}, a number of
different samples of events are selected to be used as part of the background
estimation strategy. These samples include events with
the Higgs jet candidate with a mass in the so-called ``high sideband'' (HSB) region
with 145~\gev\ < $m_J$ < 200~\gev, as well as events in a
``low sideband'' (LSB) region
with 50~\gev\ < $m_J$ < 75~\gev. In addition, 
``0-tag'' samples, with Higgs candidates that pass these various mass requirements but have no
associated $b$-tagged jets, are also selected. The 0-tag sideband samples are denoted by
LSB0 and HSB0. The 0-tag sample satisfying the SR mass
cut is denoted by CR0 to emphasize its use as a control region in the background
determination (see Section~\ref{sec:bkgndestimate}), rather than as an additional signal region.

\begin{figure}[hbpt]
\centering
\includegraphics[width=0.48\linewidth]{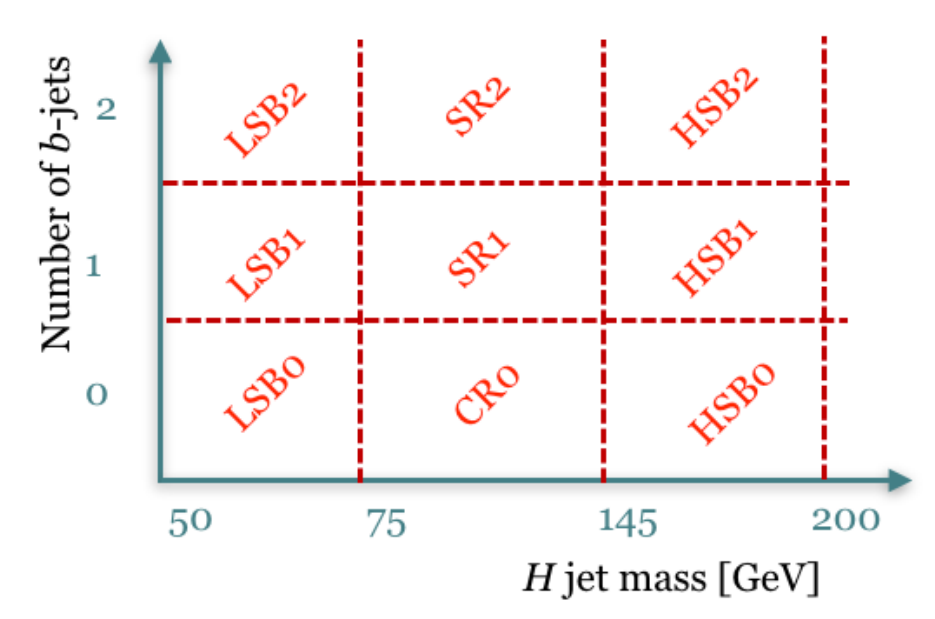}
\caption{Illustration of the definitions of the various mutually exclusive data regions.
The data are divided
into Low Sidebands (LSB), Signal Regions (SR), and High Sidebands (HSB)
according to the mass of the candidate Higgs jet. Each region is
subdivided further according to the number of $b$-tagged jets associated with the
candidate Higgs jet. The 0-tag sample satisfying the SR mass
cut is denoted by CR0 to emphasize its use as a control region in the background
determination, rather than as an additional signal region.
}\label{fig:strategy}
\end{figure}

If both large-$R$ jets fulfil the Higgs mass requirement, 
the jet with the highest number of associated $b$-tagged jets
is chosen as the Higgs candidate. In case both have  the same 
number of $b$-tagged jets, the ambiguity is resolved by assigning
the leading \pt\ large-$R$ jet
as the Higgs candidate.
 Studies of signal samples with $m(X) = 110$~\gev\ demonstrate that, in the lower $m(Y)$ region, 
  the probability to correctly identify the $H$ and $X$ jets is 98.8\%. This value
  decreases to 89.5\% for 
  $m(Y) = 4$~\tev, due to the falling efficiency in the very high mass region to tag two 
  separate $b$-tagged jets
  within the $H$ jet.

The large-$R$ jet which is not chosen as the Higgs candidate is assigned to
the $X$ hypothesis. The jet substructure variable $D_2$, described previously, 
is used to check whether the $X$ candidate jet is compatible with the two-prong structure
expected due to its assumed $X \to q\bar q'$ decay. The requirement on the
value of $D_2$ corresponds to that defined in 
Refs.~\cite{PERF-2015-03,ATL-PHYS-PUB-2015-033} to provide a constant efficiency 
of 50\% for selecting hadronic $V$ decays, when applied
along with an associated requirement that $m_J$ lies within a window
around the $V$ mass. Given that the $X$ mass is a priori unknown, the event selection does not make any requirement on the mass of the $X$ candidate jet beyond the $m_J > 50$~\gev\ restriction applied to all large-$R$ jets, however the data are interpreted in windows of $m_X$.  

The mass of the $Y \to XH$ candidate is required to be larger than 1~\tev.
The efficiencies for the 1-tag and 2-tag signal region
selections are presented as a function of $m(Y)$ and $m(X)$ in Fig.~\ref{fig:allsigeff}  for
the various signal samples. 
No signal samples with $m(Y) > 4$~\tev\ were simulated, due to the low expected sensitivity for
such high masses 
with the current data sample. However, data events with higher reconstructed 
$Y$ mass values are included in the overflow bin in the mass distributions. 

\begin{figure}[hbpt]
\centering
\subfigure[]{\includegraphics[width=0.48\linewidth]{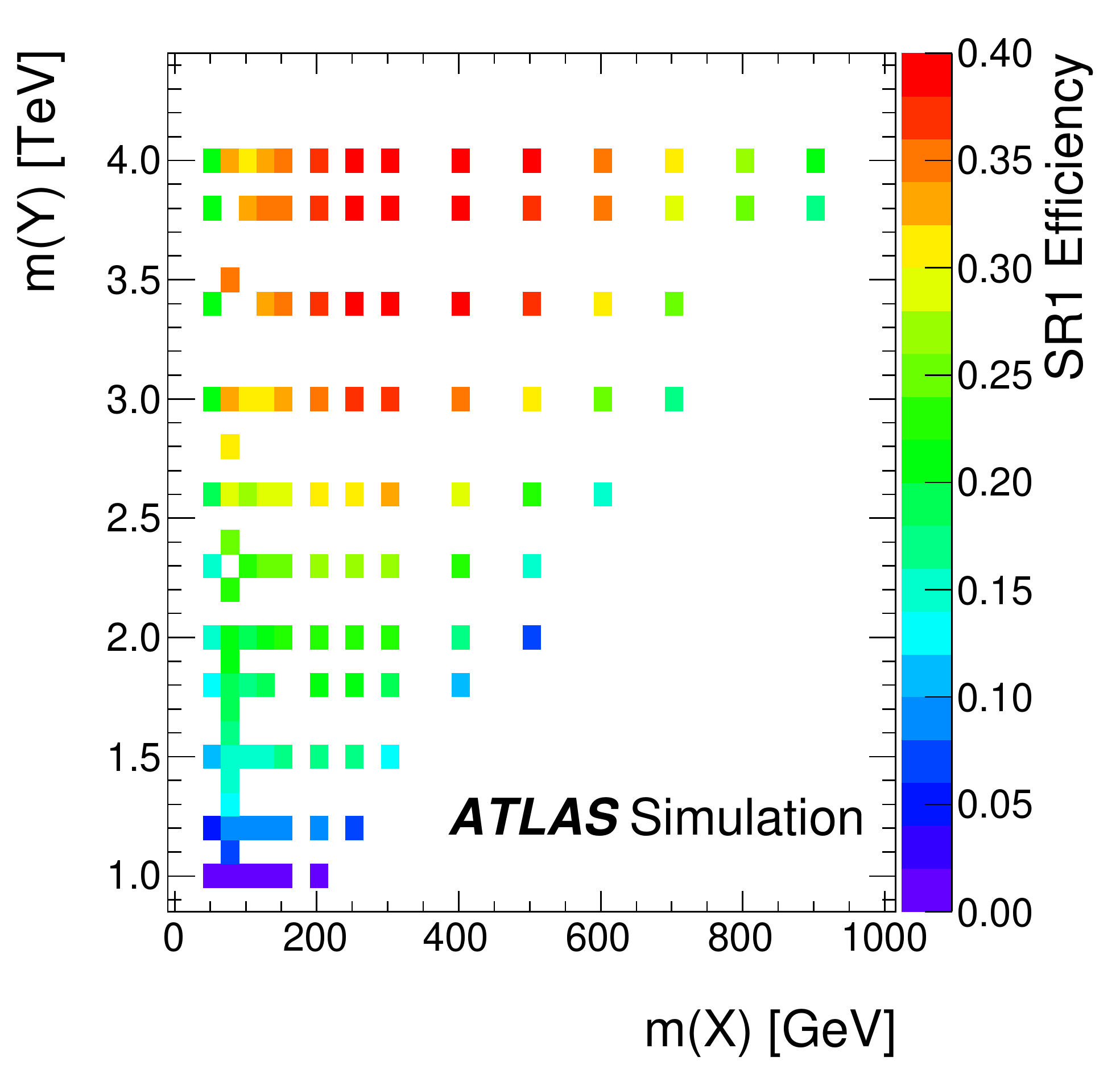}}
\subfigure[]{\includegraphics[width=0.48\linewidth]{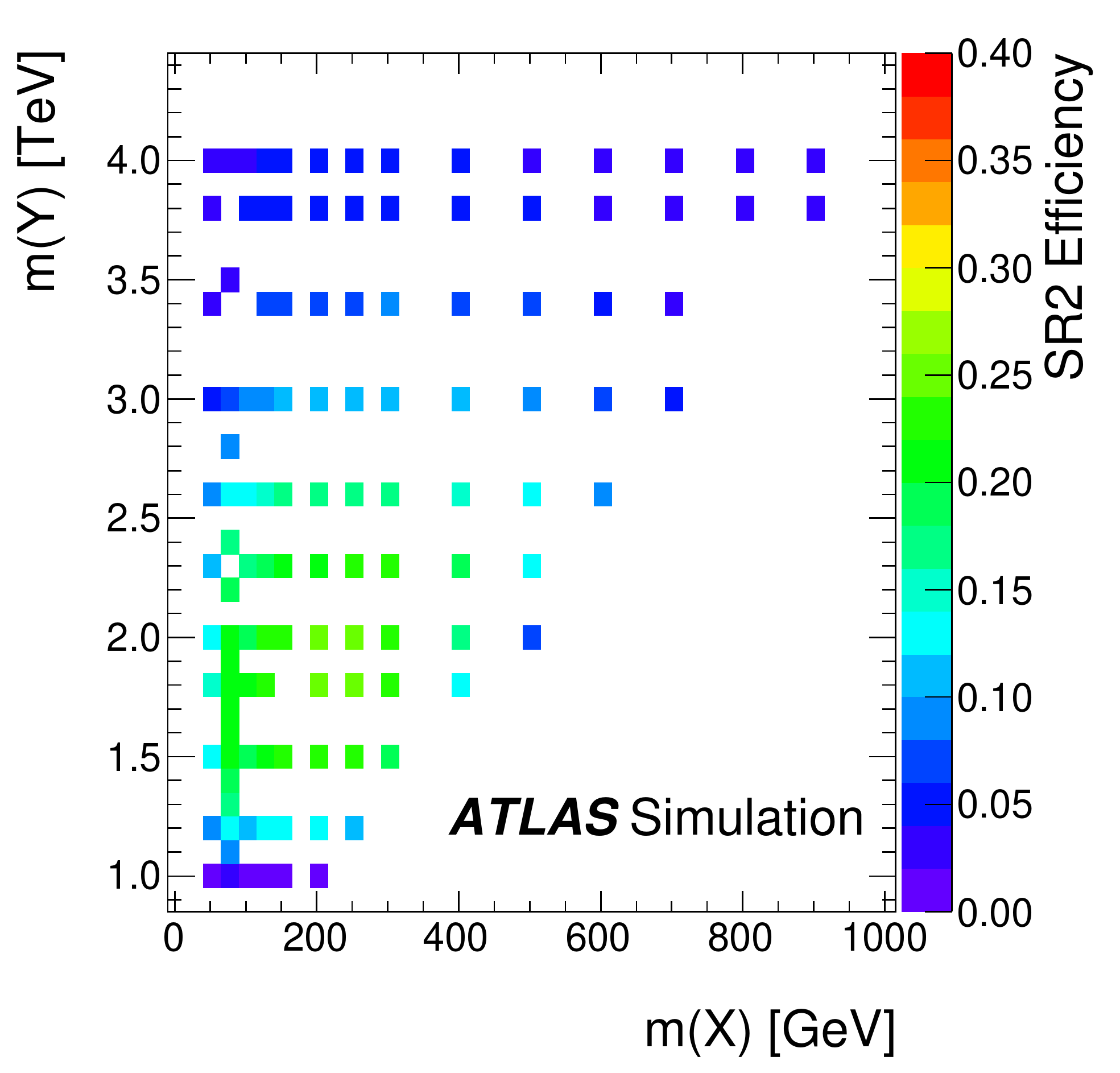}}
\caption{Efficiencies, as a function of $m(Y)$ and $m(X)$ for all the generated signal samples, for
the selection defining 
(a) the 1-tag signal region (SR1) and (b) the 2-tag signal region (SR2). 
}\label{fig:allsigeff}
\end{figure}

%-------------------------------------------------------------------------------

%-------------------------------------------------------------------------------
\section{Signal modelling}
\label{sec:signalmodel}

The signal search in the two-dimensional space of $m(Y)$ versus $m(X)$ 
employs a sliding-window technique to the $m_J$ spectrum of the
$X$ candidate jet, dividing the data into a
series of overlapping $m_J$ ranges. 
As described in Section~\ref{sec:statint}, 
for each $m_J$ window of the $X$ candidate jet, the corresponding 
$m_{JJ}$ distribution is examined for evidence of an excess due to a signal,
where $m_{JJ}$ denotes the invariant mass of the system formed by the
$H$ and $X$ candidate large-$R$ jets.

The reconstructed $X$ mass resolution varies from 12~\gev\ to 40~\gev.
The widths of the overlapping $m_J$ windows for the $X$ candidate jet 
are chosen to be around twice
the $X$ mass resolution, but also take into account the limited number of data events. 
 The central values of neighbouring $m_J$ windows are
shifted by roughly half of the $X$ mass resolution, ensuring that they overlap
and do not leave gaps in the search coverage. Studies in which simulated signals
were injected at various mass values were used to demonstrate the
reasonableness of the window choices.

The reconstructed $Y \to XH$ mass resolution varies from about 65~\gev\ to 100~\gev.
For each $m_J$ window of the $X$ candidate jet, the binning for the corresponding $m_{JJ}$
distribution is chosen to be at least as large as the $Y$ mass resolution,
dependent on the size of the data sample. 

Given the large number of possible $m(Y)$ and $m(X)$ values in the
ranges considered, a parameterization of the signal is used to
interpolate between the mass values for which signal
samples were generated. Probability distribution functions for the 
$m_{JJ}$ and $m_J$ shapes are built using the ${\tt RooKeysPdf}$ class of the 
${\tt RooFit}$ package~\cite{Verkerke:2003ir}. Multidimensional morphing methods, implemented in the ${\tt RooMomentMorphND}$ class~\cite{Baak:2014fta}, are 
subsequently applied to parameterize grid points using the surrounding four 
nearby points. The normalization of the parameterized signals is estimated 
using linear interpolation.

\section{Background Estimation}
\label{sec:bkgndestimate}

After the event selection, over 96\% of the SM background originates from multijet processes.
A data-driven 
method is used to determine the shape of the dominant multijet background,
with its normalization in the signal regions being determined with the fit
procedure described in Section~\ref{sec:statint}.
The small additional background contributions from $t\bar{t}$ and $V$+jets production
are estimated via simulation.  The background determination is performed
separately in each of the $m_J$ windows considered for the $X$ candidate jet mass.

As depicted in Fig.~\ref{fig:strategy}, the event sample, before any
specific requirements are made on the mass of the $X$ candidate jet, is separated into mutually exclusive
categories according to the mass and number of $b$-tagged jets of the Higgs boson candidate.
The SR1 and SR2 regions
are used to perform the signal search, while the other seven regions are
used as control regions to develop and validate the background
model. The multijet background component in each of the seven control 
regions is determined by subtracting from the data the 
$t\bar t$ and $V$+jets contributions predicted by simulation.

The modelling of the dominant multijet background in the SR1 and SR2 signal regions 
starts from the CR0 sample, for which the candidate Higgs jet satisfies the SR requirement on the
jet mass but has no associated b-tagged jets. The CR0 sample 
should contain negligible signal contamination. 
The basic strategy is to use events in the CR0 sample with at least 
one (two) track-jet(s) associated with the Higgs jet to model the shape of the
multijet background in the SR1 (SR2) signal region. 
However, sources of possible differences 
between the multijet components of CR0 versus SR1 and SR2
include changes in the underlying event populations due to the absence or
presence of $b$-quarks
as well as kinematic differences arising 
from the application of $b$-tagging, since the $b$-tagging efficiency depends on the
\pt\ and $\eta$ of the track-jet.   The corrections required to take these
differences into account
are extracted from the HSB events. In each of the HSB regions with different number of b-tags (HSB0, HSB1, HSB2), a pair of two-dimensional histograms is filled, one of \pt\ versus $\eta$ 
for the leading track-jet, and the other for the subleading track-jet.
Subsequently, leading and subleading track 
jet reweighting maps are created by dividing the HSB1 and HSB2 histograms
by the corresponding HSB0 histogram. A Gaussian kernel, similar to 
that applied in Ref.~\cite{PERF-2012-01}, is used to smooth 
out statistical fluctuations by taking a weighted sum of neighbouring bins. 
The weight is inversely proportional to the width of the Gaussian kernel and depends on the 
statistical uncertainty of each bin. The result is a reduction of statistical 
fluctuations, while adding negligible bias to the distributions.

To describe the SR multijet background shapes, the CR0 events are reweighted using the maps in bins of track-jet \pt\ and $\eta$ extracted from the sideband. The reweighting is performed only for the 2-tag samples, as the modelling of the shape of the multijet background in the 1-tag sideband regions without reweighting is observed to be adequate. Due to the correlations between the kinematic variables of the two leading track-jets used for the reweighting, multiple reweighting iterations are performed, until the track-jet properties, \pt\ and $\eta$, are matched within statistical uncertainties. 

The background modelling is validated by using the same method to
predict the background in the LSB regions in the low
mass sidebands, and then comparing with the data and obtaining good agreement. Agreement is also confirmed 
in the signal region by integrating over all values
of the mass of the $X$ candidate jet, thereby diluting  any possible signal contamination 
to a negligible level.

%-------------------------------------------------------------------------------

%-------------------------------------------------------------------------------
\section{Systematic uncertainties}
\label{sec:syst}
The main systematic uncertainty in the background estimate arises from the potential 
mis-modelling of the background processes.  As described in Section~\ref{sec:bkgndestimate},
the multijet background shapes in the two signal regions are estimated directly 
from data using sideband regions. To determine
the systematic uncertainty in this method, the multijet background shapes
in the HSB1 and HSB2 regions are compared,  in each $m_J$ window 
of the $X$ candidate jet, to those of the
reweighted HSB0 multijet distributions. The differences, in bins of $m_{JJ}$, fitted with a line are used as the multijet modelling systematic uncertainties. The largest observed shape difference is found to be approximately 12\%, while the smallest is 3\%. A normalization uncertainty
of 30\% is assigned to the small \ttbar  background,   
based on the ATLAS \ttbar differential cross-section measurement~\cite{TOPQ-2014-15}.
The same 
uncertainty is conservatively applied to the small $V$+jets background component.

The uncertainty in the combined 2015+2016 integrated luminosity is 2.1\%. 
It is derived, following a methodology similar to that detailed in Ref.~\cite{DAPR-2013-01}, 
from a calibration of the luminosity scale using $x$--$y$ beam--separation 
scans performed in August 2015 and May 2016. This uncertainty is 
applied to the yields of both the signal and the small $t\bar t$ and 
$V+$jets backgrounds, which are not determined from data.

Additional systematic uncertainties in the signal acceptance arise from the choice of PDF 
set and the uncertainty in the amount of initial- and final-state radiation. For all the two-dimensional 
mass grid points, a constant 5\% uncertainty is applied and covers both effects.

Uncertainties related to the signal parameterization method are estimated by comparing the 
mass distributions of generated signal points to those predicted from morphing nearby 
points in the two-dimensional space of $m(Y)$ versus $m(X)$. 
The largest observed normalization deviation is $\sim$8\%, while differences in
the signal shape also reach levels up to $\sim$8\%. Both effects are included as uniform uncorrelated systematic uncertainties.

Systematic uncertainties which account for experimental
effects on the signal shape include the large-$R$ jet mass 
scale and mass resolution, as well as $D_2$ and the $b$-tagging.  
The impact of each of these effects is evaluated by shifting each variable according to these systematic uncertainties and then re-performing
the signal parameterization.
Based on the measurements
documented in Refs.~\cite{PERF-2012-02,ATL-PHYS-PUB-2015-033}, 
a 2\% uncertainty is assigned for the large-$R$ jets \pt scale, while uncertainties of
20\% and 15\%, respectively, are assigned for the large-$R$ jet mass 
resolution and $D_2$ resolution.  
Uncertainties in the correction factors for the $b$-tagging 
are applied to the simulated event samples by looking at dedicated 
flavour-enriched samples in data. An additional term is included to 
extrapolate the measured uncertainties to the high-\pT region of interest. 
This term is calculated from simulated events by considering variations 
of the quantities affecting the $b$-tagging performance such as the impact 
parameter resolution, percentage of poorly measured tracks, description 
of the detector material, and track multiplicity per jet. The dominant effect 
on the uncertainty when extrapolating to high \pT is related to the change
in tagging efficiency when smearing the track impact parameters 
based on the resolution measured in data and simulation. The 
uncertainty in the $b$-tagging efficiency is measured as a function of $b$-jet \pt 
and $\eta$ for track-jets with \pt < 250~\gev\, while for higher \pt\ values
it is extrapolated using simulation~\cite{PERF-2012-04}. These uncertainties vary between 2\% and 8\% 
in the lower \pt\ range, and rise to approximately 9\% for track-jets with \pt\ > 400~\gev.

%-------------------------------------------------------------------------------

%-------------------------------------------------------------------------------
\section{Statistical treatment and results}
\label{sec:statint}
The signal search in the two-dimensional space of $m(Y)$ versus $m(X)$ is performed by applying a sliding-window technique to the $m_J$ spectrum of the $X$ candidate jet, which divides the data into a series of overlapping $m_J$ ranges. For each $m_J$ window,
a binned maximum-likelihood fit is then performed  to 
the resultant $m_{JJ}$ distributions in both SR1 and SR2 signal regions simultaneously. 
A test statistic based 
on the profile likelihood ratio~\cite{Cowan:2010js} is used to test hypothesized values 
of the global signal strength ($\mu$), corresponding to the HVT model. The systematic uncertainties, described
previously, are modelled with Gaussian or log-normal constraint terms 
(nuisance parameters) in the definition of the likelihood function.

Each $m_J$ window of the $X$ candidate jet is fitted independently, but as the mass ranges overlap, the  fit results are correlated. In
each fit, the normalizations of the multijet backgrounds in the SR1 and SR2 
signal regions are treated as free parameters. All of  the systematic uncertainties are 
treated as correlated between the two signal regions. 

The data distributions and fit results are shown for three example
$m_J$  windows of the $X$ candidate jet in Fig.~\ref{fig:expostfit0}--\ref{fig:expostfit2}. 
In each case, the left (right) upper plot shows the
$m_{JJ}$ distribution in the SR1 (SR2) signal region, and the results of the simultaneous fit are superimposed on the plots.
Examples of possible signal models are also shown, with arbitrary overall normalization, illustrating
the corresponding contributions that would be expected in the SR1 and SR2 regions.
The data distributions are well described by the fitted
SM background, and there is no significant evidence of a signal. The same conclusion is valid for all $m_J$ windows
of the $X$ candidate jet.

\begin{figure}[hbpt]
\centering
\subfigure[][\label{fig:expostfit00}]{\includegraphics[width=0.48\hsize]{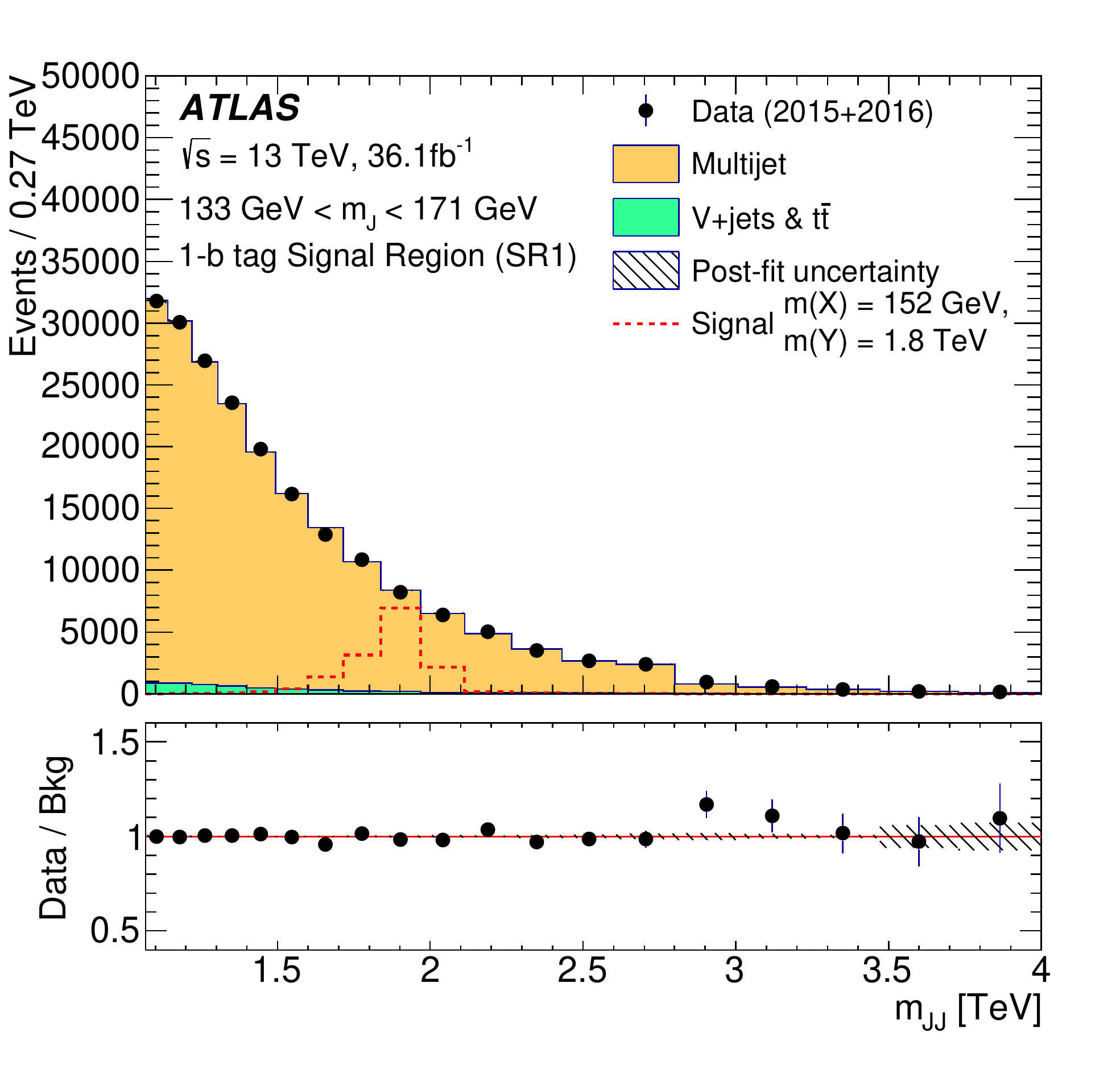}}
\subfigure[][\label{fig:expostfit01}]{\includegraphics[width=0.48\hsize]{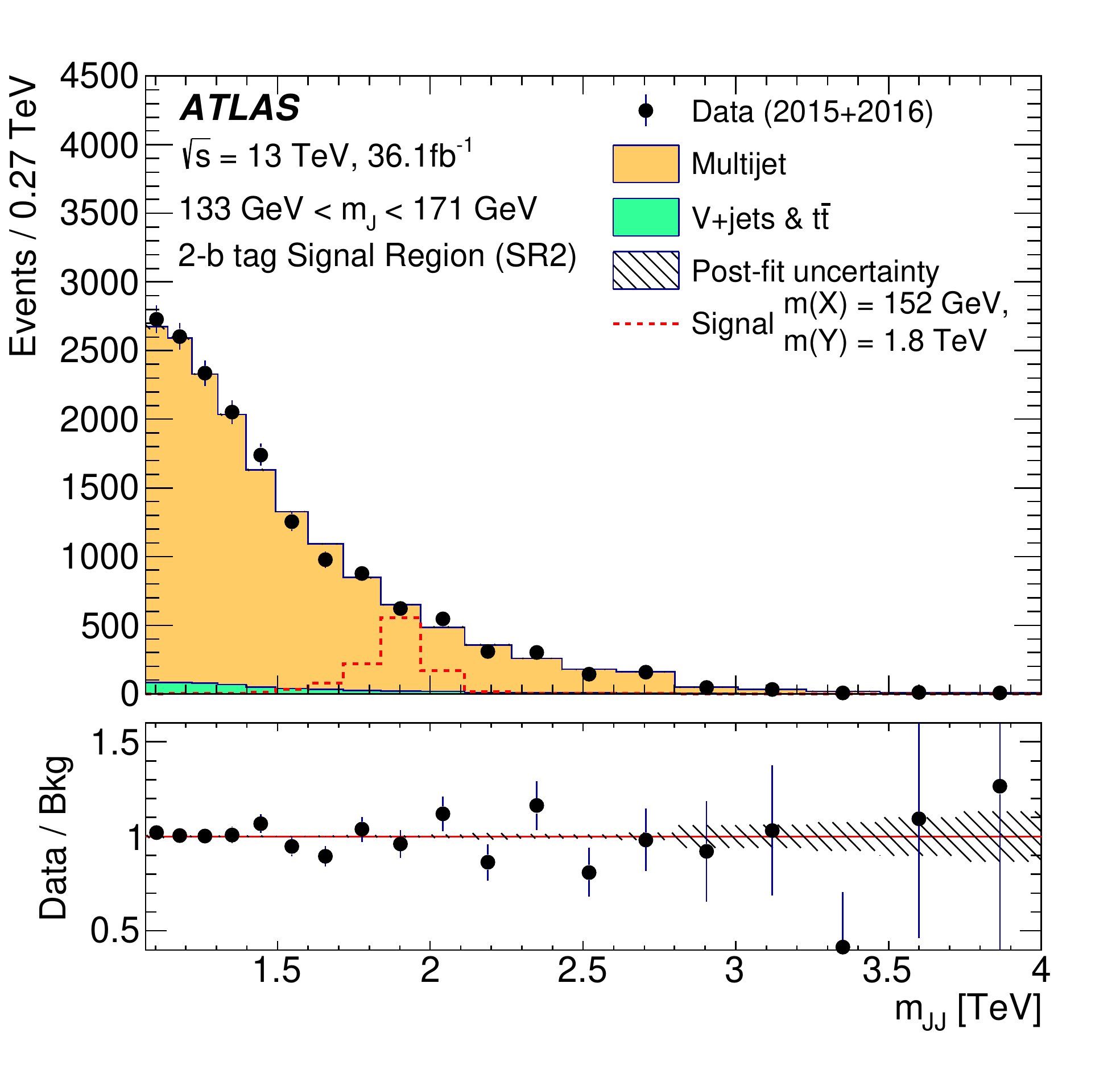}}
\subfigure[][\label{fig:expostfit01l}]{\includegraphics[width=0.48\hsize]{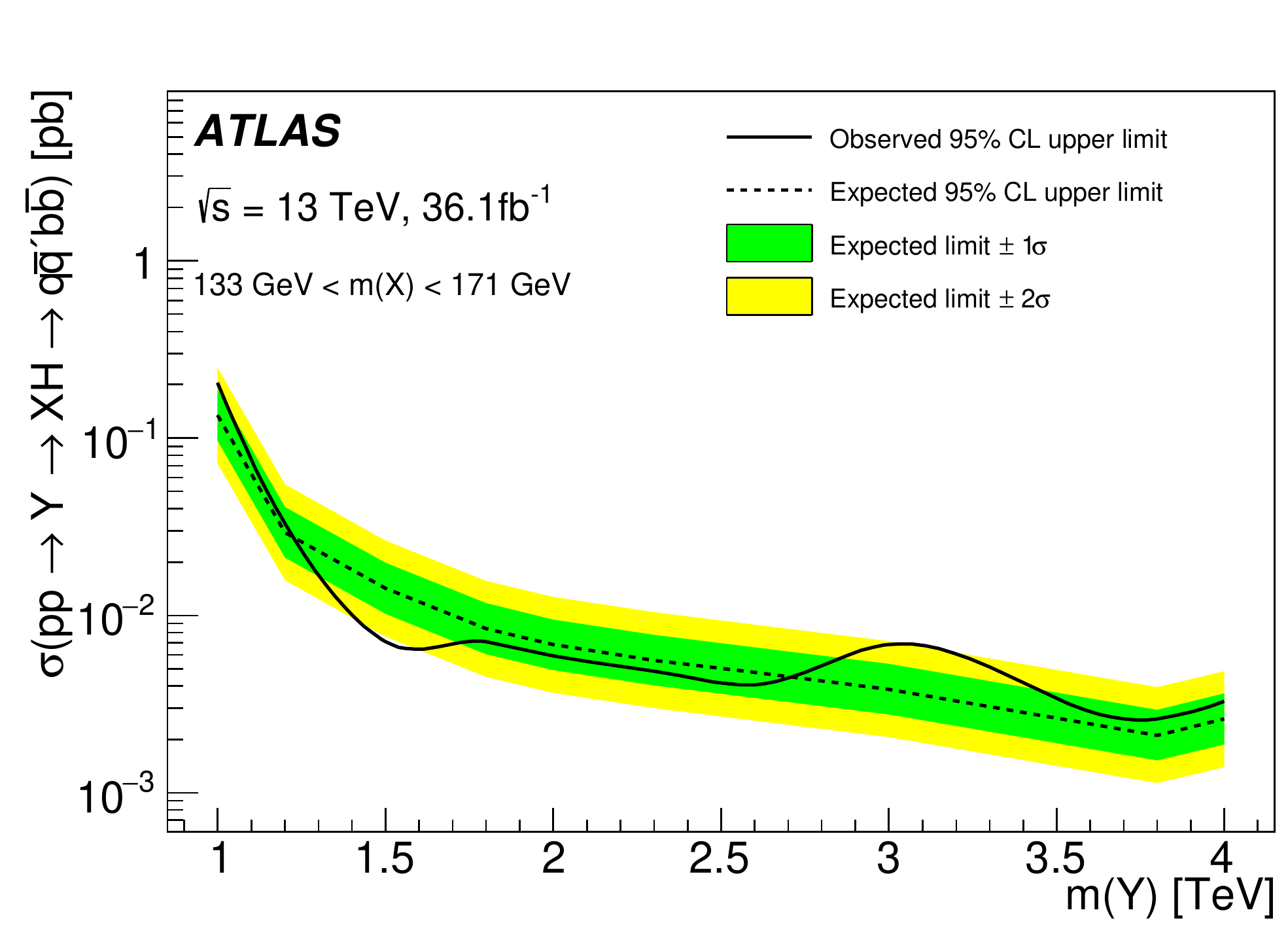}}
\caption{The $m_{JJ}$ mass distributions in the $m_J$ window of the $X$ candidate jet 
from 133~\gev\ to 171~\gev\ after the 
likelihood fit for events in (a) the 1-tag signal region (SR1) and (b) the 2-tag signal region (SR2).  The
highest mass bin includes any overflows.
The background expectation is given by the filled histograms and the ratio of the observed data to 
the background (Data/Bkg) is shown in the lower panel. 
The uncertainties shown are those after the fit described in the text.  
One particular example of a possible
signal model, namely with $m(X) = $ 152~\gev\ and $m(Y) = $ 1.8~\tev,
is overlaid with an arbitrary overall normalization, illustrating
the corresponding contributions that would be expected in the SR1 and SR2 regions.
Panel (c) shows the resultant 95\% CL upper 
limit on the production cross-section, $\sigma(pp \to Y \to XH \to q\bar q'b\bar b)$.} \label{fig:expostfit0}
\end{figure}

\begin{figure}[hbpt]
\centering
\subfigure[][\label{fig:expostfit10}]{\includegraphics[width=0.48\hsize]{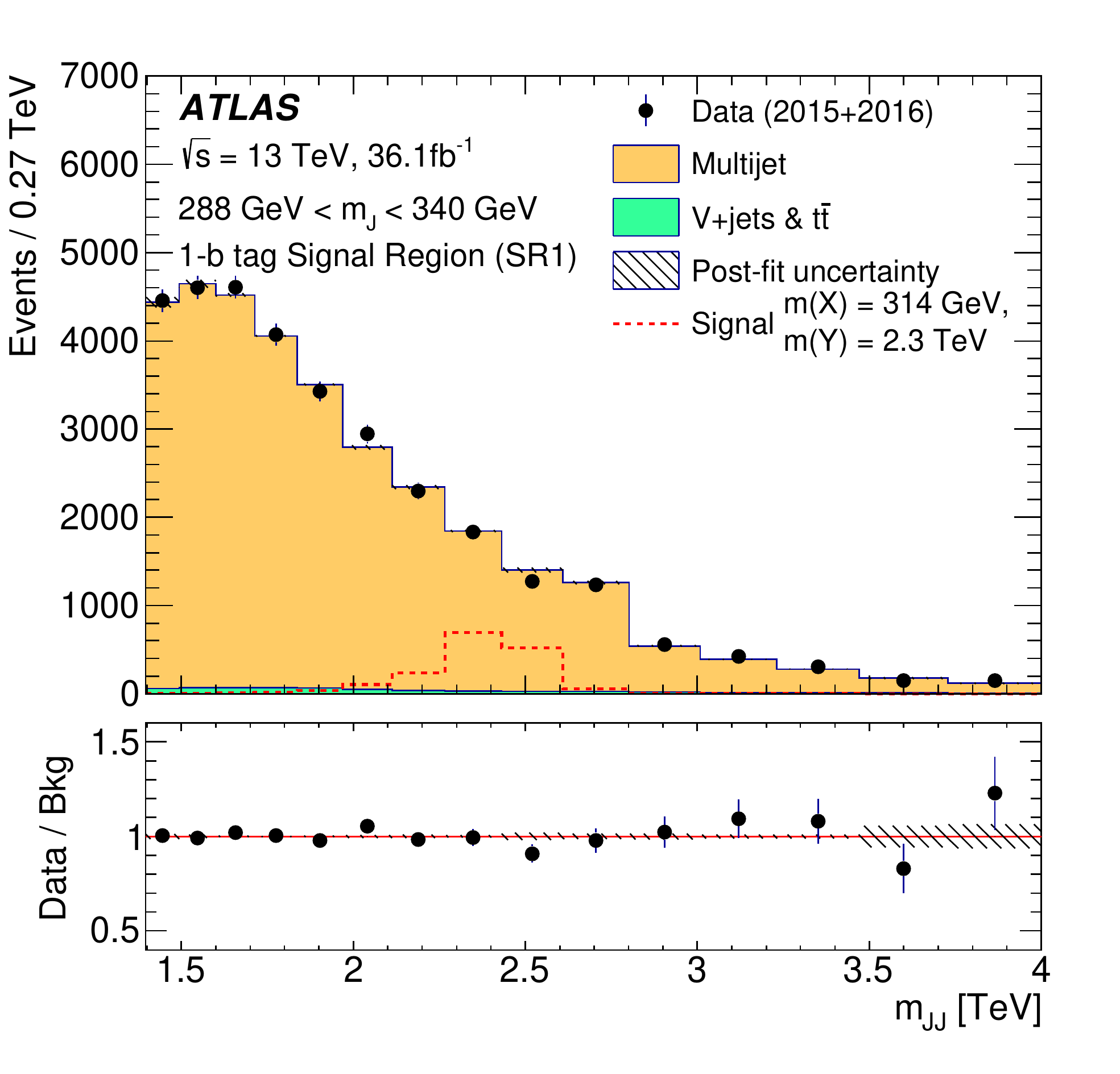}}
\subfigure[][\label{fig:expostfit11}]{\includegraphics[width=0.48\hsize]{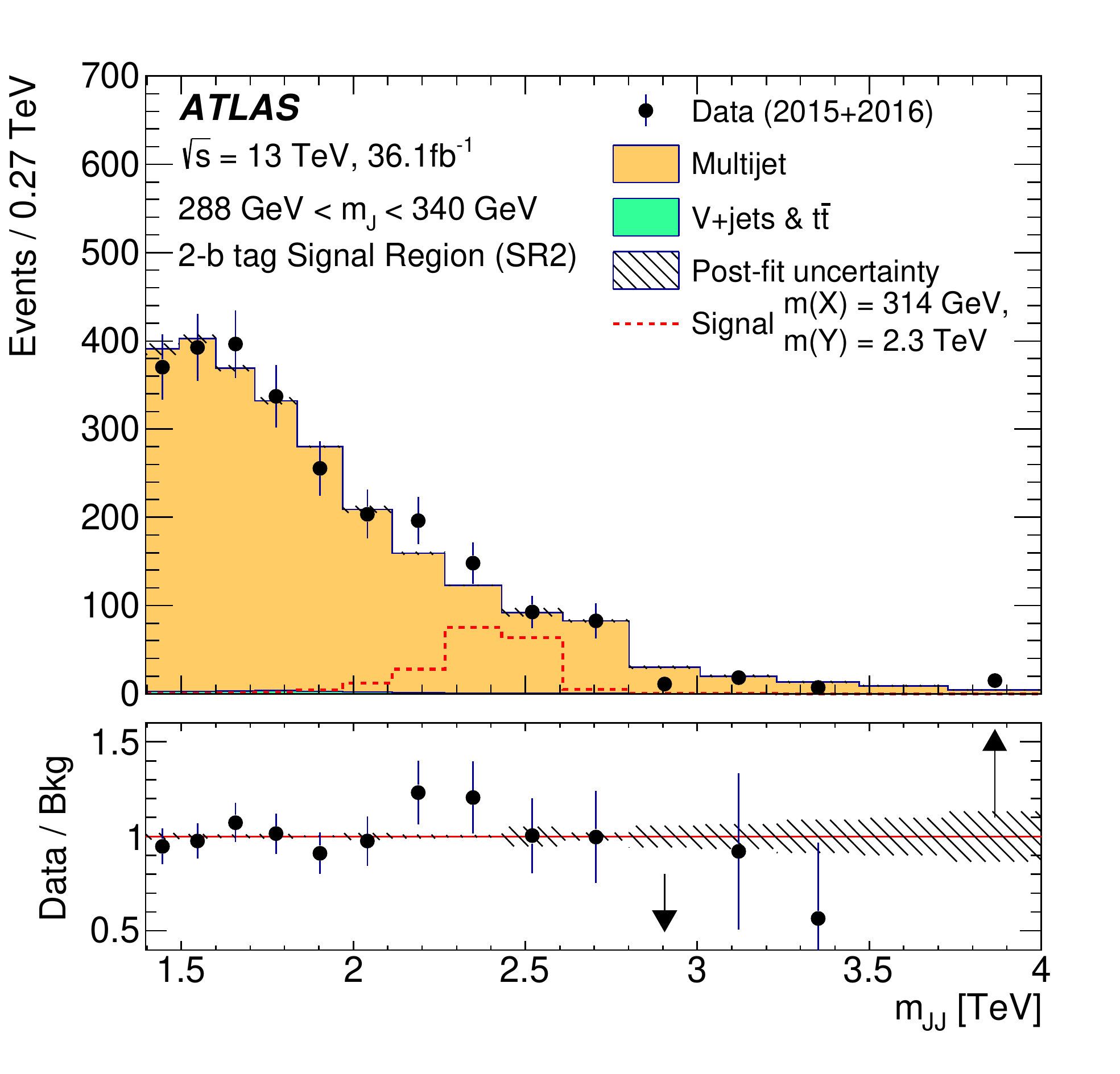}}
\subfigure[][\label{fig:expostfit11l}]{\includegraphics[width=0.48\hsize]{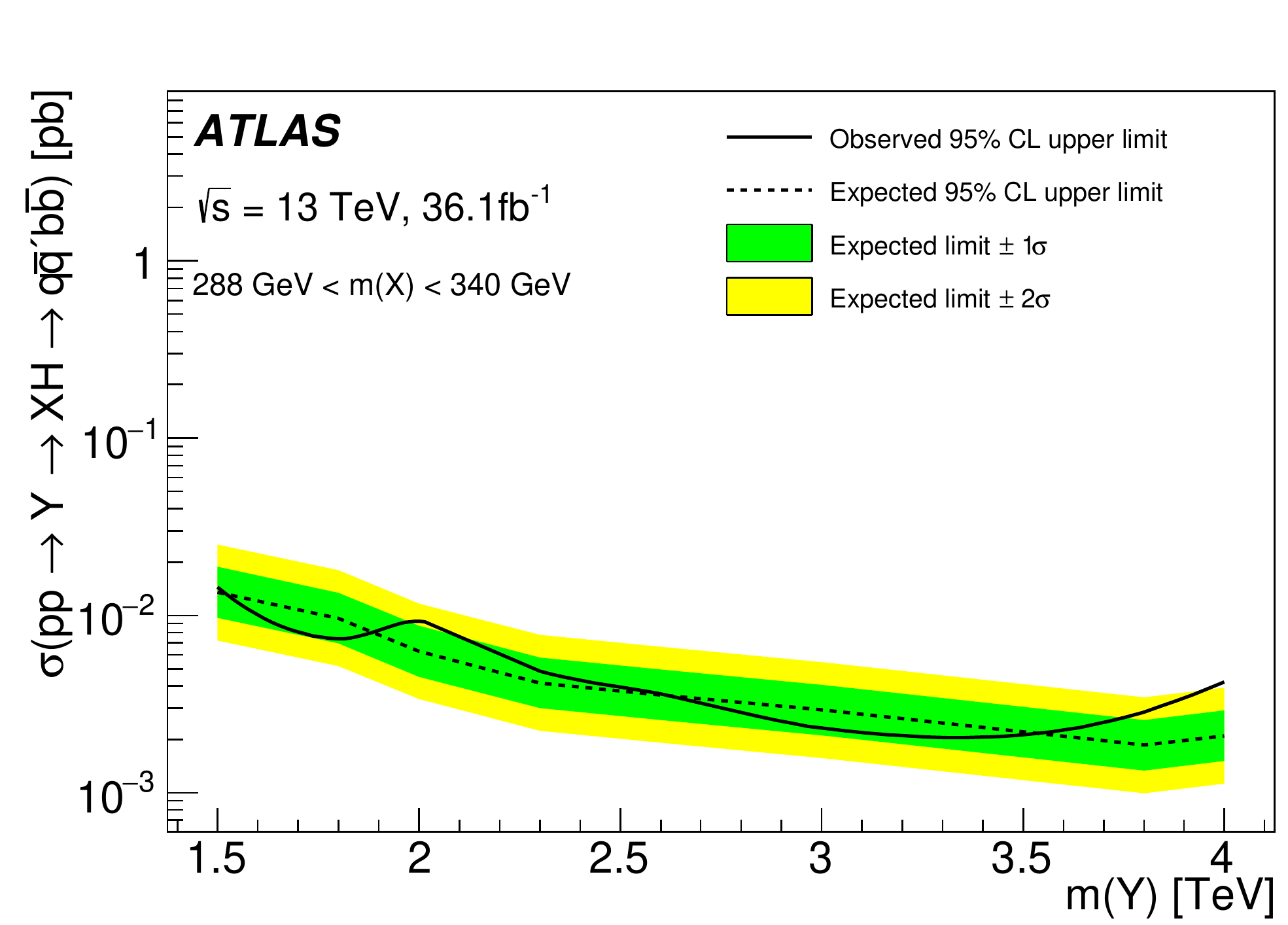}}
\caption{The $m_{JJ}$ mass distributions in the $m_J$ window of the $X$ candidate jet 
 from 288~\gev\ to 340~\gev\ after the 
likelihood fit for events in (a) the 1-tag signal region (SR1) and (b) the 2-tag signal region (SR2). 
The highest mass bin includes any overflows.
The background expectation is given by the filled histograms and the ratio of the 
observed data to the background (Data/Bkg) is shown in the lower panel. 
The uncertainties shown are those after the fit described in the text. 
One particular example of a possible
signal model, namely with $m(X) = $ 314~\gev\ and $m(Y) = $ 2.3~\tev,
is overlaid with an arbitrary overall normalization, illustrating
the corresponding contributions that would be expected in the SR1 and SR2 regions.
Panel (c) shows the resultant 95\% CL upper 
limit on the production cross-section, $\sigma(pp \to Y \to XH \to q\bar q'b\bar b)$. } \label{fig:expostfit1}
\end{figure}

\begin{figure}[hbpt]
\centering
\subfigure[][\label{fig:expostfit20}]{\includegraphics[width=0.48\hsize]{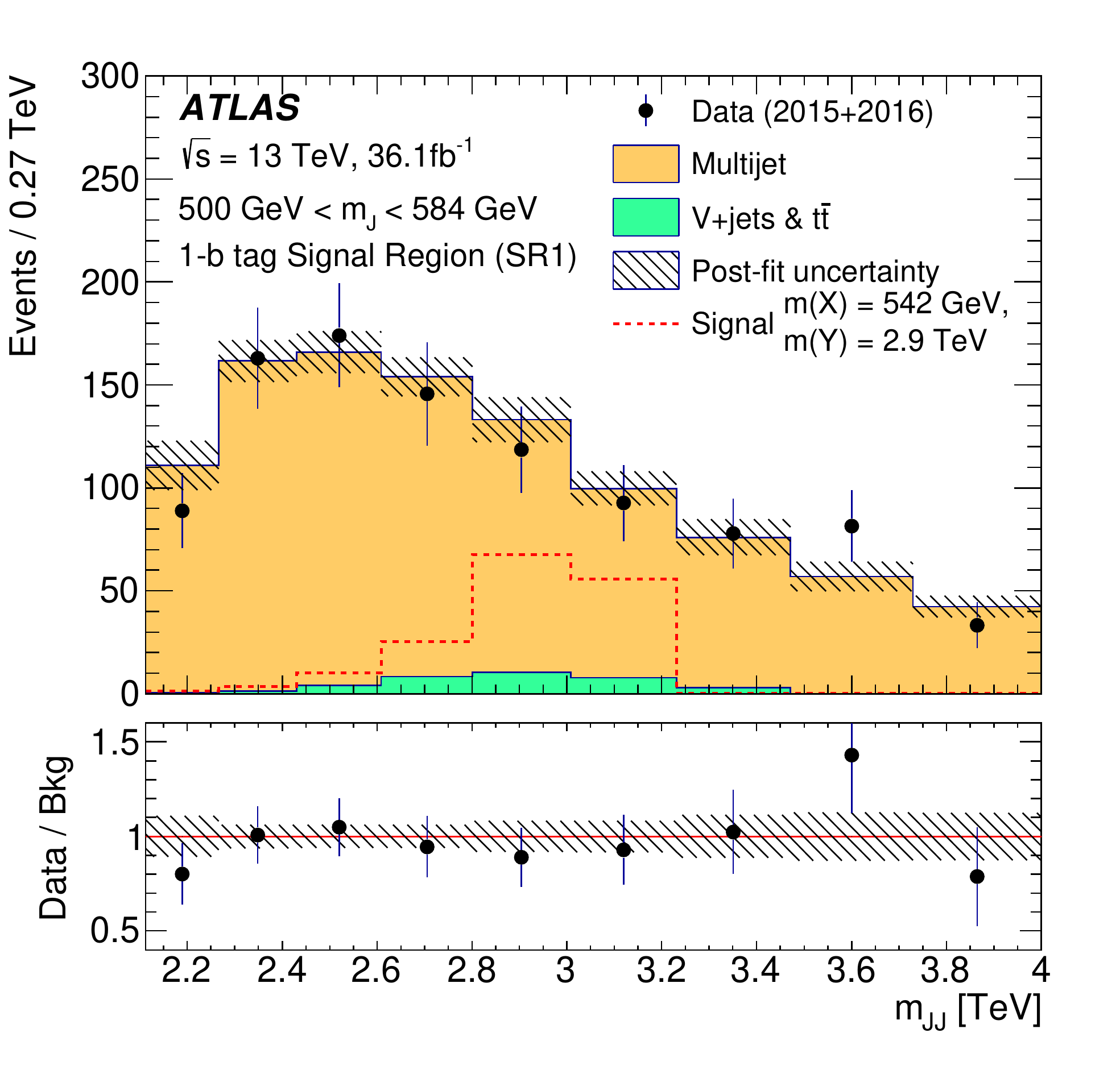}}
\subfigure[][\label{fig:expostfit21}]{\includegraphics[width=0.48\hsize]{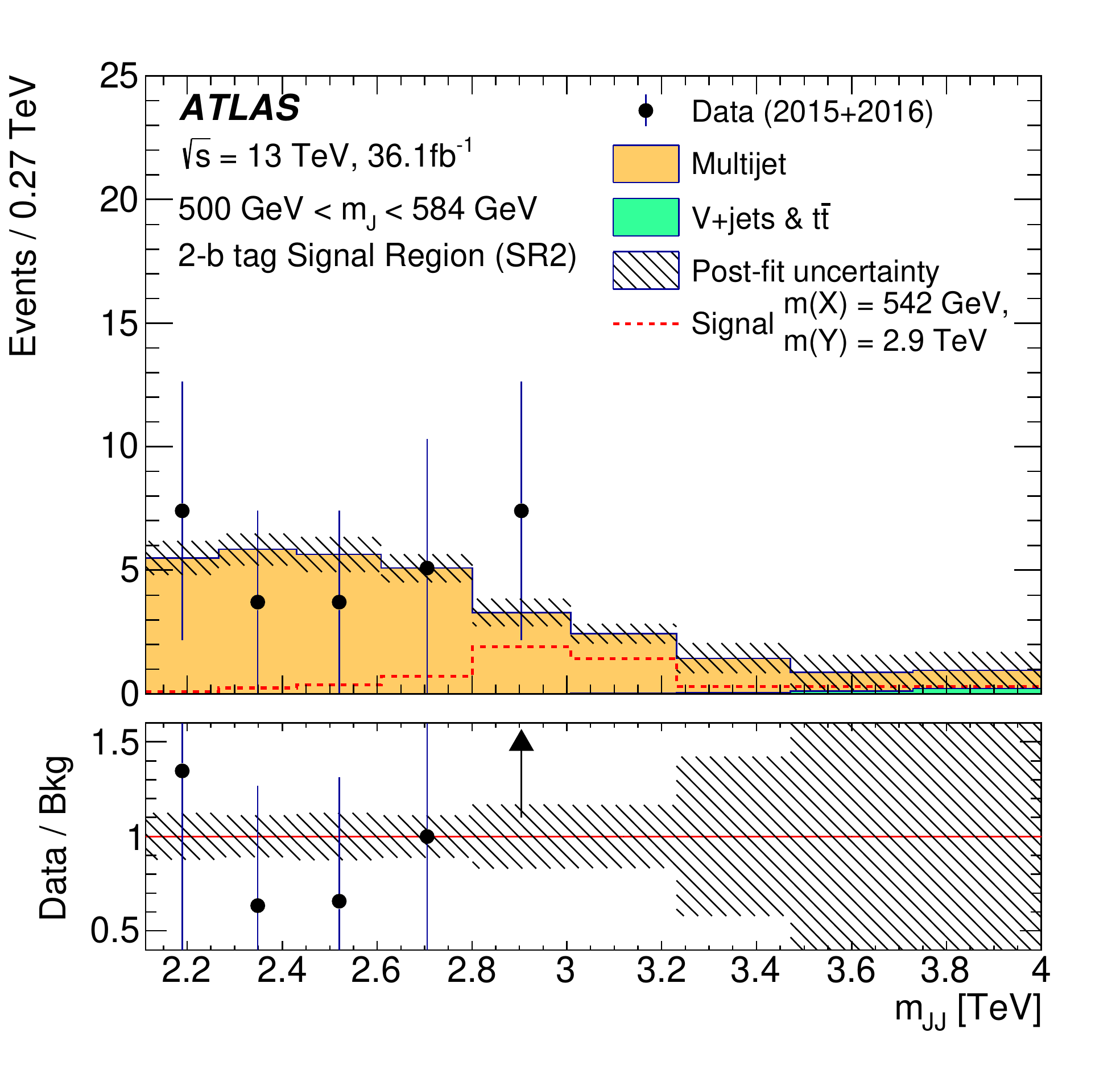}}
\subfigure[][\label{fig:expostfit21l}]{\includegraphics[width=0.48\hsize]{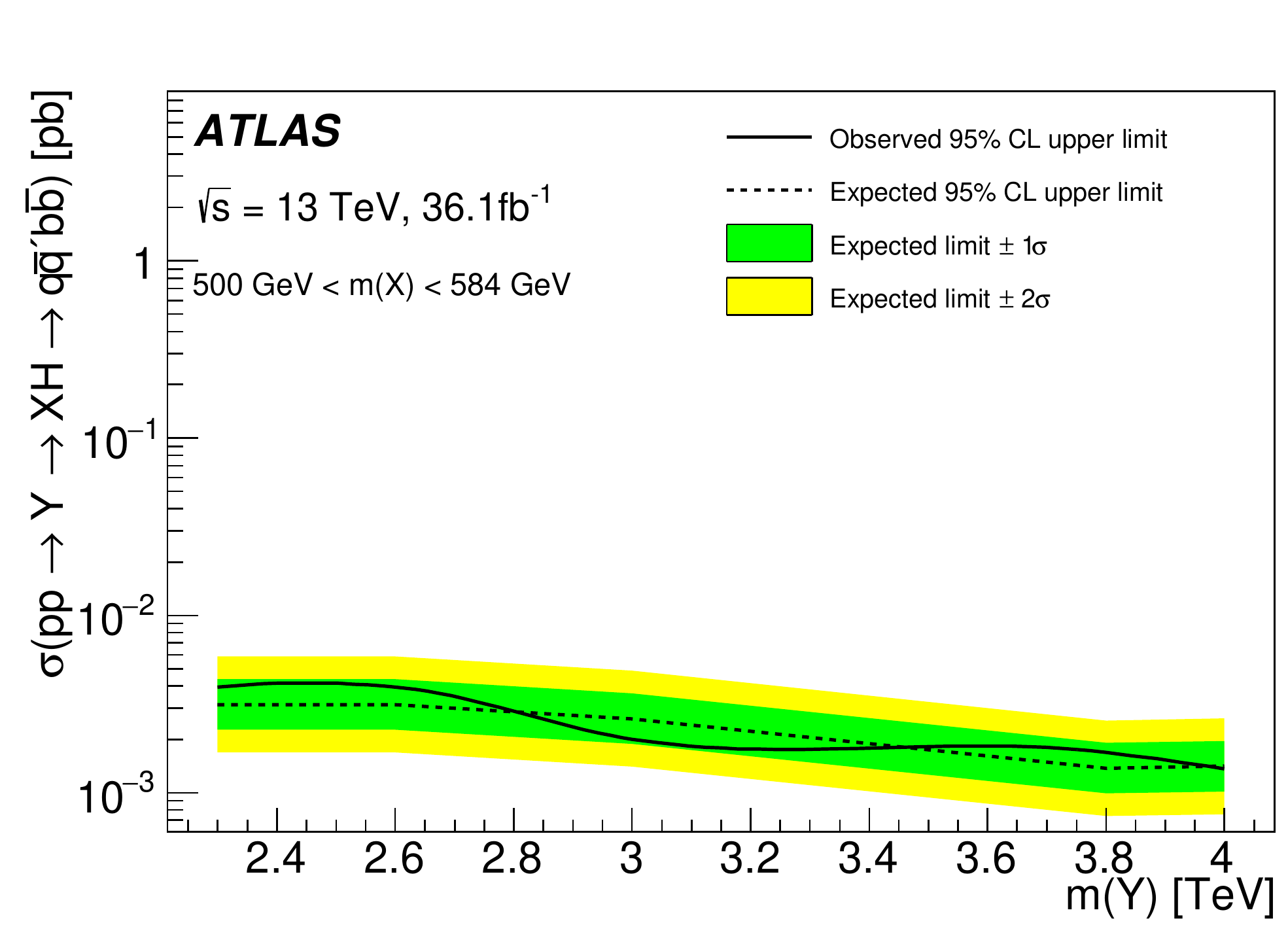}}
\caption{The $m_{JJ}$ mass distributions in the $m_J$ window of the $X$ candidate jet 
 from 500~\gev\ to 584~\gev\ 
after the likelihood fit for events in (a) the 1-tag signal region (SR1) and (b) the 2-tag signal region (SR2).
The highest mass bin includes any overflows.
The background expectation is given by the filled histograms and the ratio of the 
observed data to the background (Data/Bkg) is shown in the lower panel. 
The uncertainties shown are those after the fit described in the text. 
One particular example of a possible
signal model, namely with 
$m(X) = $ 542~\gev\ and $m(Y) = $ 2.9~\tev,
is overlaid with an arbitrary overall normalization, illustrating
the corresponding contributions that would be expected in the SR1 and SR2 regions.
Panel (c) shows the resultant 95\% CL upper 
limit on the production cross-section, $\sigma(pp \to Y \to XH \to q\bar q'b\bar b)$. } \label{fig:expostfit2}
\end{figure}

The fits to the data are used to set upper limits on the production cross-section,
 $\sigma(pp \to Y \to XH \to q\bar q'b\bar b)$. 
Exclusion limits are computed using the CL$_{\text{s}}$ method~\cite{Read:2002hq}, with a 
value of $\mu$ regarded as excluded at the 95\% confidence level (CL) when CL$_{\text{s}}$ is 
less than 5\%. The corresponding limit plots are included in panel (c) of 
Fig.~\ref{fig:expostfit0}--\ref{fig:expostfit2}.

A summary of the expected and observed 95\% CL limits on the 
production cross-section, $\sigma(pp \to Y \to XH \to q\bar q'b\bar b)$, is 
given in the two-dimensional plane of $m(Y)$ versus $m(X)$ 
in Fig.~\ref{fig:limits2D}. For $X$ mass windows around the $V$ and Higgs boson mass values, 
the results are found to be compatible with the dedicated ATLAS $VH$ and $HH$ searches reported 
in Refs.~\cite{Aaboud:2017ahz} and \cite{EXOT-2015-11}, respectively. 
To visualize both the magnitude and the sign of discrepancies between
the expected and observed limits, Fig.~\ref{fig:deviation2D} 
presents the differences between the expected and observed limits, expressed in multiples
of $\sigma$, the equivalent Gaussian significance. 
The maximum observed deviation has a local significance
of about 2.5$\sigma$, corresponding to a global deviation of 1.2$\sigma$. 

\begin{figure}[hbpt]
\centering
\subfigure[][\label{fig:limits2D0}]{\includegraphics[width=0.48\linewidth]{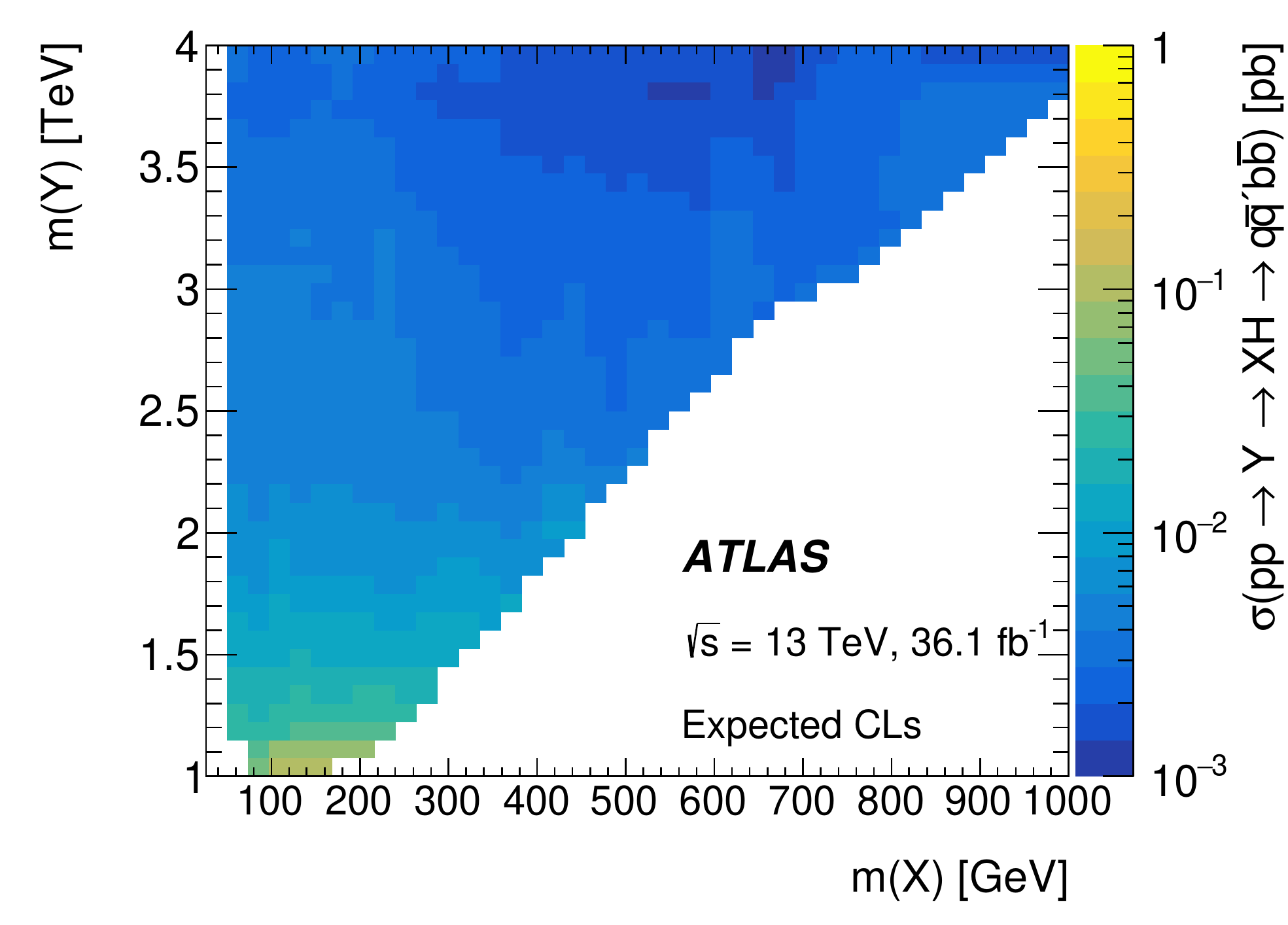}}
\subfigure[][\label{fig:limits2D1}]{\includegraphics[width=0.48\linewidth]{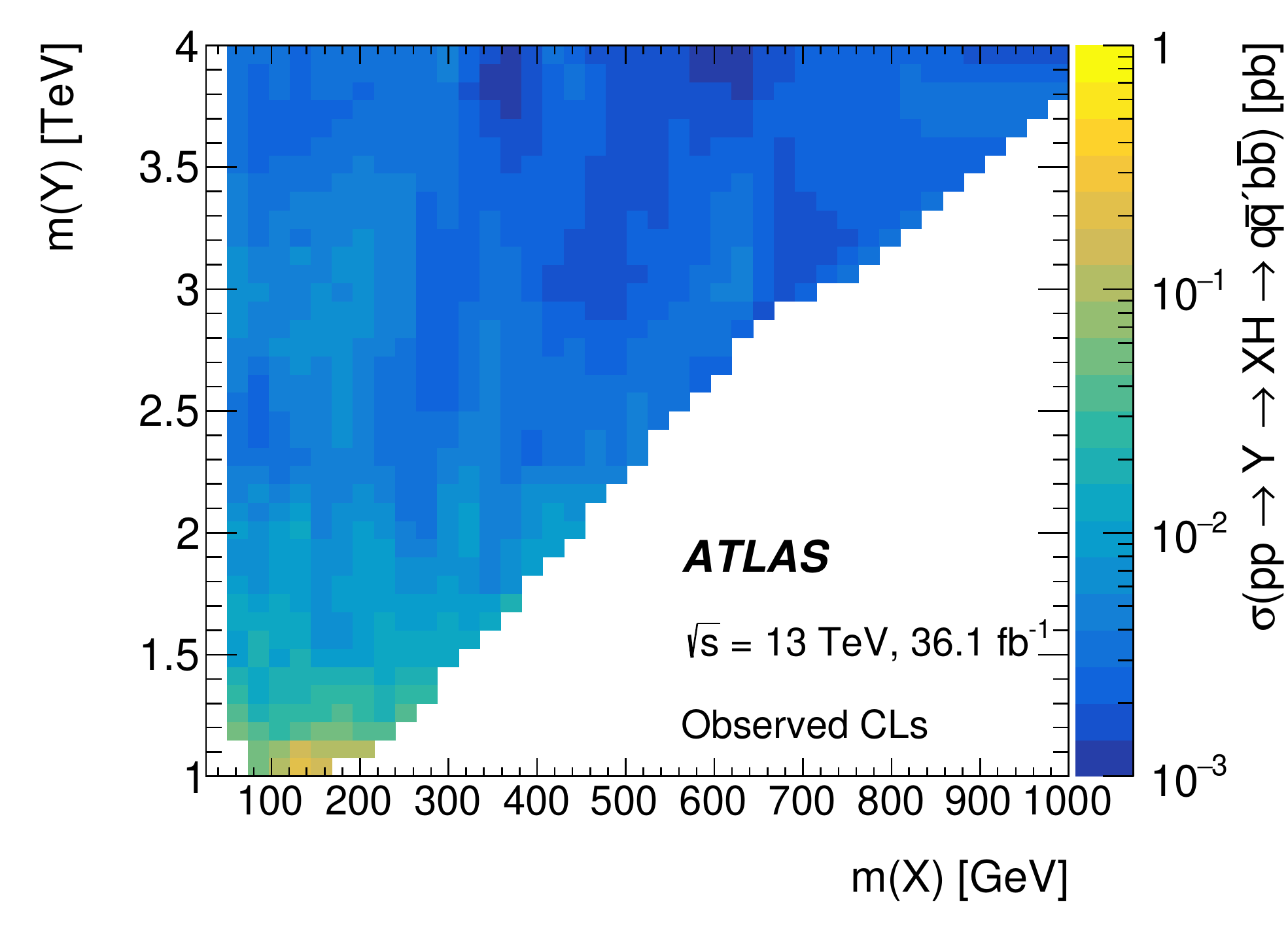}}
\caption{The (a) expected and (b) observed 95\% CL limits on the production cross-section 
$\sigma(pp \to Y \to XH \to q\bar q'b\bar b)$
(z-axis) in the two-dimensional space of $m(Y)$ versus $m(X)$.}
\label{fig:limits2D}
\end{figure}

\begin{figure}[hbpt]
\centering
\includegraphics[width=0.48\linewidth]{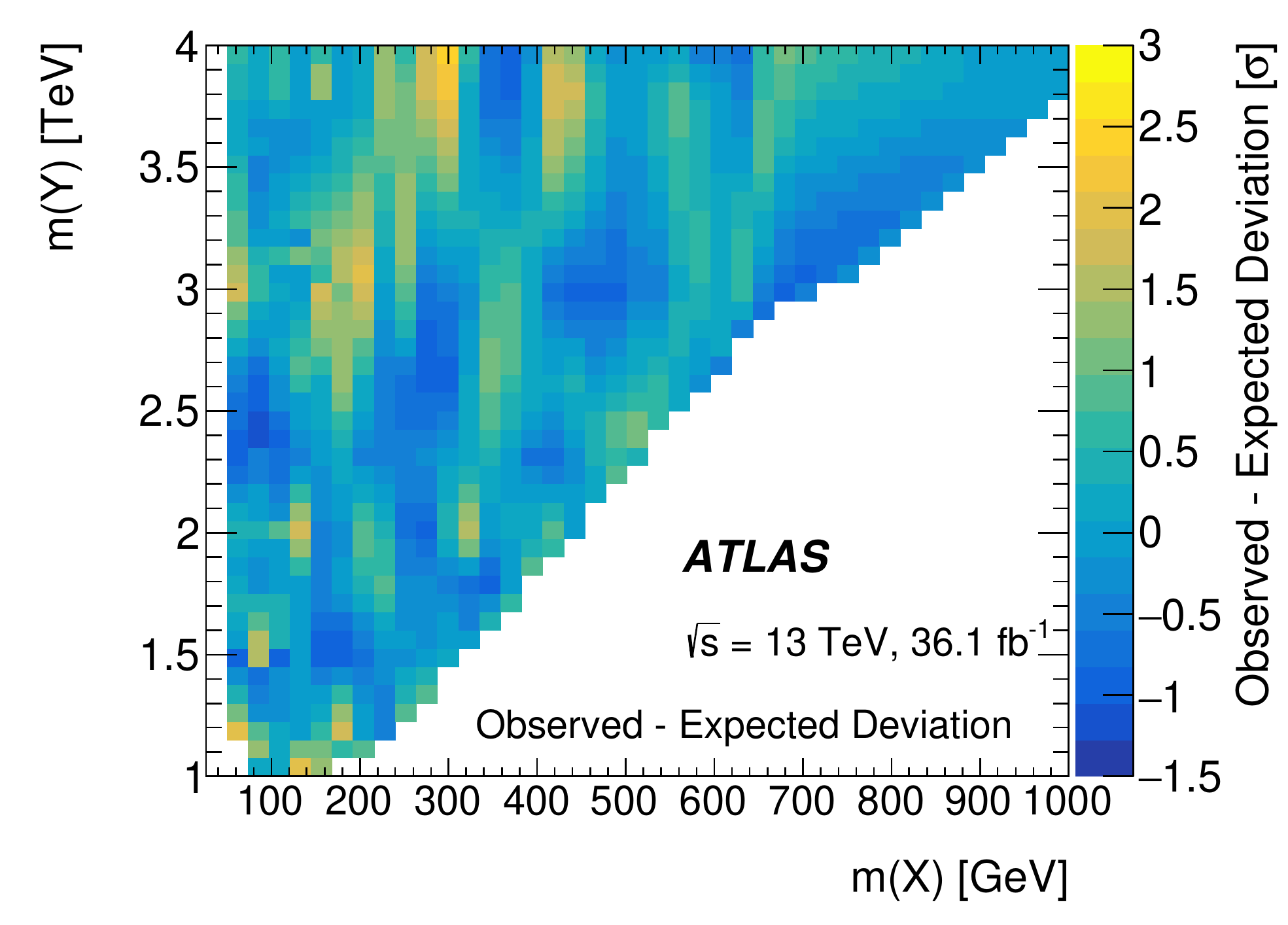}
\caption{Differences between the observed and the expected 95\% CL limits on the 
resonance production cross-section
$\sigma(pp \to Y \to XH \to q\bar q'b\bar b)$, expressed in terms of multiples of 
$\sigma$, the equivalent Gaussian significance. }
\label{fig:deviation2D}
\end{figure}

%-------------------------------------------------------------------------------

%-------------------------------------------------------------------------------
\FloatBarrier
\section{Conclusion}
\label{sec:conclusion}

A search for new heavy resonances $Y$ decaying into a Higgs boson and a new particle $X$ was carried out using 36.1~\ifb~ of $pp$ collision data collected at $\sqrt{s} =$ 13 \tev~ by 
ATLAS during the 2015 and 2016 runs of the CERN Large Hadron Collider. 
The $Y \to XH \to q\bar q'b\bar{b}$ channel was studied using a generic approach in the topological regime where both bosons are highly Lorentz-boosted, 
and each is reconstructed as a single jet with a large radius parameter. 
Jet substructure and $b$-tagging techniques are exploited to tag the $X$ and Higgs 
bosons and to reduce the dominant multijet background. Values of $Y$ ($X$)
mass in the range of 1~\tev\ to 4~\tev\ (50~\gev\ to 1000~\gev) were considered.

A search for evidence of an excess in the $XH$ mass spectrum was made in a large number of overlapping sliding windows in the mass of the $X$ particle.
The data are found to be in 
agreement with the Standard Model background expectations and only small deviations are observed, with local (global) significance of no more than 2.5$\sigma$
(1.2$\sigma$).  Within the framework of a modified Heavy Vector Triplet model,
upper limits on the resonance
production cross-section $\sigma(pp \to Y \to XH \to q\bar q'b\bar b)$ were set at 95\% CL in the two-dimensional space of the 
$Y$ mass versus the $X$ mass.

%-------------------------------------------------------------------------------

%-------------------------------------------------------------------------------
\section*{Acknowledgments}
%-------------------------------------------------------------------------------
% Acknowledgements for papers with collision data
% Version 16-Nov-2017

% Standard acknowledgements start here
%----------------------------------------------
We thank CERN for the very successful operation of the LHC, as well as the
support staff from our institutions without whom ATLAS could not be
operated efficiently.

We acknowledge the support of ANPCyT, Argentina; YerPhI, Armenia; ARC, Australia; BMWFW and FWF, Austria; ANAS, Azerbaijan; SSTC, Belarus; CNPq and FAPESP, Brazil; NSERC, NRC and CFI, Canada; CERN; CONICYT, Chile; CAS, MOST and NSFC, China; COLCIENCIAS, Colombia; MSMT CR, MPO CR and VSC CR, Czech Republic; DNRF and DNSRC, Denmark; IN2P3-CNRS, CEA-DRF/IRFU, France; SRNSF, Georgia; BMBF, HGF, and MPG, Germany; GSRT, Greece; RGC, Hong Kong SAR, China; ISF, I-CORE and Benoziyo Center, Israel; INFN, Italy; MEXT and JSPS, Japan; CNRST, Morocco; NWO, Netherlands; RCN, Norway; MNiSW and NCN, Poland; FCT, Portugal; MNE/IFA, Romania; MES of Russia and NRC KI, Russian Federation; JINR; MESTD, Serbia; MSSR, Slovakia; ARRS and MIZ\v{S}, Slovenia; DST/NRF, South Africa; MINECO, Spain; SRC and Wallenberg Foundation, Sweden; SERI, SNSF and Cantons of Bern and Geneva, Switzerland; MOST, Taiwan; TAEK, Turkey; STFC, United Kingdom; DOE and NSF, United States of America. In addition, individual groups and members have received support from BCKDF, the Canada Council, CANARIE, CRC, Compute Canada, FQRNT, and the Ontario Innovation Trust, Canada; EPLANET, ERC, ERDF, FP7, Horizon 2020 and Marie Sk{\l}odowska-Curie Actions, European Union; Investissements d'Avenir Labex and Idex, ANR, R{\'e}gion Auvergne and Fondation Partager le Savoir, France; DFG and AvH Foundation, Germany; Herakleitos, Thales and Aristeia programmes co-financed by EU-ESF and the Greek NSRF; BSF, GIF and Minerva, Israel; BRF, Norway; CERCA Programme Generalitat de Catalunya, Generalitat Valenciana, Spain; the Royal Society and Leverhulme Trust, United Kingdom.

The crucial computing support from all WLCG partners is acknowledged gratefully, in particular from CERN, the ATLAS Tier-1 facilities at TRIUMF (Canada), NDGF (Denmark, Norway, Sweden), CC-IN2P3 (France), KIT/GridKA (Germany), INFN-CNAF (Italy), NL-T1 (Netherlands), PIC (Spain), ASGC (Taiwan), RAL (UK) and BNL (USA), the Tier-2 facilities worldwide and large non-WLCG resource providers. Major contributors of computing resources are listed in Ref.~\cite{ATL-GEN-PUB-2016-002}.
%----------------------------------------------

\clearpage
\printbibliography

\newpage 
% ATLAS Collaboration author list
% Data extracted on 18-Aug-2017 for paper reference EXOT-2017-04
%\documentclass[11pt]{article}
%\usepackage{a4wide}\begin{document}
\begin{flushleft}
{\Large The ATLAS Collaboration}

\bigskip

M.~Aaboud$^\textrm{\scriptsize 137d}$,
G.~Aad$^\textrm{\scriptsize 88}$,
B.~Abbott$^\textrm{\scriptsize 115}$,
O.~Abdinov$^\textrm{\scriptsize 12}$$^{,*}$,
B.~Abeloos$^\textrm{\scriptsize 119}$,
S.H.~Abidi$^\textrm{\scriptsize 161}$,
O.S.~AbouZeid$^\textrm{\scriptsize 139}$,
N.L.~Abraham$^\textrm{\scriptsize 151}$,
H.~Abramowicz$^\textrm{\scriptsize 155}$,
H.~Abreu$^\textrm{\scriptsize 154}$
% ...
\bigskip
\\
% ...
$^{12}$ Institute of Physics, Azerbaijan Academy of Sciences, Baku, Azerbaijan\\
% ...
$^{88}$ CPPM, Aix-Marseille Universit{\'e} and CNRS/IN2P3, Marseille, France\\
% ...
$^{115}$ Homer L. Dodge Department of Physics and Astronomy, University of Oklahoma, Norman OK, United States of America\\
% ...
$^{119}$ LAL, Univ. Paris-Sud, CNRS/IN2P3, Universit{\'e} Paris-Saclay, Orsay, France\\
% ...
$^{137}$ $^{(a)}$ Facult{\'e} des Sciences Ain Chock, R{\'e}seau Universitaire de Physique des Hautes Energies - Universit{\'e} Hassan II, Casablanca; $^{(b)}$ Centre National de l'Energie des Sciences Techniques Nucleaires, Rabat; $^{(c)}$ Facult{\'e} des Sciences Semlalia, Universit{\'e} Cadi Ayyad, LPHEA-Marrakech; $^{(d)}$ Facult{\'e} des Sciences, Universit{\'e} Mohamed Premier and LPTPM, Oujda; $^{(e)}$ Facult{\'e} des sciences, Universit{\'e} Mohammed V, Rabat, Morocco\\
% ...
$^{139}$ Santa Cruz Institute for Particle Physics, University of California Santa Cruz, Santa Cruz CA, United States of America\\
% ...
$^{151}$ Department of Physics and Astronomy, University of Sussex, Brighton, United Kingdom\\
% ...
$^{154}$ Department of Physics, Technion: Israel Institute of Technology, Haifa, Israel\\
$^{155}$ Raymond and Beverly Sackler School of Physics and Astronomy, Tel Aviv University, Tel Aviv, Israel\\
% ...
$^{161}$ Department of Physics, University of Toronto, Toronto ON, Canada\\
% ...
$^{a}$ Also at Department of Physics, King's College London, London, United Kingdom\\
$^{b}$ Also at Institute of Physics, Azerbaijan Academy of Sciences, Baku, Azerbaijan\\
$^{c}$ Also at Novosibirsk State University, Novosibirsk, Russia\\
$^{d}$ Also at TRIUMF, Vancouver BC, Canada\\
$^{e}$ Also at Department of Physics {\&} Astronomy, University of Louisville, Louisville, KY, United States of America\\
% ...
$^{*}$ Deceased
\end{flushleft}

%\end{document}
% Created with xml2latex.py

%-------------------------------------------------------------------------------
%\clearpage
%\appendix
%\part*{Auxiliary material}
%\addcontentsline{toc}{part}{Auxiliary material}
%-------------------------------------------------------------------------------
%\input{sections/Appendix.tex}

\end{document}